# Topological states in partially-$\mathcal{PT}$-symmetric azimuthal potentials

Yaroslav V. Kartashov,[1,2] Vladimir V. Konotop,[3] and Lluis Torner[1]

[1]ICFO-Institut de Ciencies Fotoniques, and Universitat Politecnica de Catalunya, Mediterranean Technology Park, 08860 Castelldefels (Barcelona), Spain
[2]Institute of Spectroscopy, Russian Academy of Sciences, Troitsk, Moscow Region, 142190, Russia
[3]Centro de Física Teórica e Computacional and Departamento de Física, Faculdade de Ciências, Universidade de Lisboa, Campo Grande 2, Edifício C8, Lisboa 1749-016, Portugal

We introduce partially-parity-time ($p\mathcal{PT}$)-symmetric azimuthal potentials composed from individual $\mathcal{PT}$-symmetric cells located on a ring, where two azimuthal directions are *nonequivalent* in a sense that in such potentials excitations carrying topological dislocations exhibit different dynamics for different directions of energy circulation in the initial field distribution. Such non-conservative ratchet-like structures support rich families of *stable* vortex solitons in cubic nonlinear media, whose properties depend on the *sign* of the topological charge due to the nonequivalence of azimuthal directions. In contrast, oppositely charged vortex solitons remain equivalent in similar fully $\mathcal{PT}$-symmetric potentials. The vortex solitons in the $p\mathcal{PT}$- and $\mathcal{PT}$-symmetric potentials are shown to feature qualitatively different internal current distributions, which are described by different discrete rotation symmetries of the intensity profiles.

The evolution of nonlinear waves carrying topological phase dislocations is a physical problem of fundamental importance attracting attention in various areas of physics, including optics, matter waves, hydrodynamics, cavities and electron beams. Such waves are of particular interest because of the salient role they play in numerous classical and quantum phenomena [1,2]. The propagation of vortex-carrying beams is especially intriguing in the presence of transverse modulation of the refractive index of the material [3]. *Conservative* potentials arising due to such modulations play a strong stabilizing role for vortex states in nonlinear media. They arrest collapse for two-dimensional beams in cubic media and suppress azimuthal modulation instabilities of bright vortex solitons that are ubiquitous in uniform media with focusing nonlinearity [4,5]. The intensity distributions of vortex solitons in inhomogeneous media reflect the symmetry of the underlying potentials [6,7]. While in some potentials stable vortex solitons may maintain simple ring-like shapes [8], they become strongly modulated in conventional periodic systems, such as square [9,10], hexagonal [11], and honeycomb [12] optical lattices, and in photonic crystals. The discrete rotation symmetry of such potentials imposes restrictions on the available topological charges of most compact symmetric vortex states [13], which may not hold for extended excitations [14]. The common feature of such conservative potentials is the *equivalence* of two azimuthal directions, manifested in the identical parameters of vortex solitons with opposite topological charges.

A class of *non-conservative* parity-time ($\mathcal{PT}$)-symmetric potentials has a remarkable feature: the transition from a purely real to a complex eigenvalue spectrum, referred to as $\mathcal{PT}$-symmetry breaking, occurs at a critical depth of the imaginary part of the structure [15]. Such transition manifests itself in the qualitative modification of wave evolution. Optical guiding structures with transversally symmetric refractive index and anti-symmetric gain/loss landscapes, provide unique platforms for the exploration of the effects associated with the $\mathcal{PT}$-symmetry breaking, since they allow realization of potentials where the $\mathcal{PT}$-symmetry condition $\mathcal{R}(\mathbf{r})=\mathcal{R}^*(-\mathbf{r})$ holds [16]. The breakup of $\mathcal{PT}$-symmetry in localized linear potentials was demonstrated in [17], the concept is extended to periodic structures [18] and nonlinear states in such systems as isolated $\mathcal{PT}$-symmetric waveguides [19], nondispersive and dispersive couplers [20], oligomers [21], discrete arrays [22,23], as well as to continuous nonlinear [24], linear [19,25], and mixed [26] lattices. The spectrum of the complex potential may remain real even if the potential is not $\mathcal{PT}$-symmetric [27] in one-dimensional or only partially $\mathcal{PT}$-symmetric [28] in multidimensional problems.

In contrast to *conservative* systems, where vortex-free modes do not feature internal currents, in $\mathcal{PT}$-*symmetric* potentials such currents are necessary to achieve stable beam propagation. Thus, although $\mathcal{PT}$-symmetric potentials may support stationary states with symmetric intensity distributions, there exists a certain *selected* direction in them, defined by the local currents. This phenomenon is responsible for removal of degeneracy of vortices in discrete arrays with an embedded $\mathcal{PT}$-symmetric defect [23]. In this Letter we use this property to construct *partially-$\mathcal{PT}$-symmetric continuous azimuthal potentials* from fully $\mathcal{PT}$-symmetric cells placed on a ring, where azimuthal directions become *nonequivalent*. We address vortex solitons in such structures (for discrete arrays see [23,29]) and show that due to the *nonequivalence* of two azimuthal directions, the properties of vortex solitons depend not only on the absolute value of their topological charge, but also on its *sign*.

We consider the evolution of paraxial beams in a focusing cubic medium with simultaneous transverse modulation of the refractive index and of the gain and losses, that is described by the dimensionless Schrödinger equation for the field amplitude $q$:

$$i\frac{\partial q}{\partial \xi} = -\frac{1}{2}\nabla^2 q - [p_{\text{re}}\mathcal{R}_{\text{re}}(\eta,\zeta) - ip_{\text{im}}\mathcal{R}_{\text{im}}(\eta,\zeta)]q - q|q|^2. \tag{1}$$

Here $\nabla^2 = \partial^2/\partial\eta^2 + \partial^2/\partial\zeta^2$ is the Laplacian, $\eta,\zeta$ are the transverse coordinates, $\xi$ is the normalized propagation distance, the depths of the real $p_{\text{re}}$ and imaginary $p_{\text{im}}$ parts of the complex potential $\mathcal{R} = p_{\text{re}}\mathcal{R}_{\text{re}} - ip_{\text{im}}\mathcal{R}_{\text{im}}$ are determined by the complex refractive index profile $\delta n_{\text{re}} - i\delta n_{\text{im}}$, where $\delta n_{\text{re}} \sim p_{\text{re}}\mathcal{R}_{\text{re}}$ and $\delta n_{\text{im}} \sim p_{\text{im}}\mathcal{R}_{\text{im}}$. We build a complex azimuthal potential (see [30] for a conservative counterparts) by placing $N$ Gaussian waveguides that *individually respect* $\mathcal{PT}$-symmetry, equidistantly on a ring of the radius $\rho$:

$$\begin{aligned}\mathcal{R}_{\text{re}} &= \sum_{k=1}^{N} e^{-[(\eta-\rho\cos\phi_k)^2+(\zeta-\rho\sin\phi_k)^2]/a^2}, \\ \mathcal{R}_{\text{im}} &= \sum_{k=1}^{N} \sigma^{k-1}(\zeta\cos\phi_k - \eta\sin\phi_k)e^{-[(\eta-\rho\cos\phi_k)^2+(\zeta-\rho\sin\phi_k)^2]/a^2},\end{aligned} \tag{2}$$

where $\sigma = \pm 1$, $\phi_k = 2\pi(k-1)/N$, and $a$ is the waveguide width. When $\sigma = -1$ the potential is $\mathcal{PT}$-symmetric, i.e. $\mathcal{R}(\eta,\zeta) = \mathcal{R}(-\eta,\zeta) = \mathcal{R}^*(\eta,-\zeta) = \mathcal{R}^*(-\eta,-\zeta)$ meaning that the potential is even along the horizontal direction and $\mathcal{PT}$-symmetric along the vertical direction, while for $\sigma = +1$ it is still $\mathcal{PT}$-symmetric vertically, but the invariance under the horizontal inversion is replaced by the invariance under simultaneous inversion of $\eta$ and $\zeta$: $\mathcal{R}(\eta,\zeta) = \mathcal{R}^*(-\eta,\zeta) = \mathcal{R}^*(\eta,-\zeta) \neq \mathcal{R}^*(-\eta,-\zeta)$ (Fig. 1). In the latter case, the potential is referred as partially-$\mathcal{PT}$ ($p\mathcal{PT}$)–symmetric [28]. Such potentials have different orders of discrete rotation symmetries. We consider even values of $N$ for which the $\mathcal{PT}$-symmetric potential belongs to the $C_{N/2,v}$ point group of $N/2$ rotations by the angles $\phi_{\sigma=-1} = 4\pi/N$ and $N/2$ mirror reflections, while $p\mathcal{PT}$-symmetric potentials belong to the $C_N$ group of $N$ rotations by the angles $\phi_{\sigma=+1} = 2\pi/N$. It is convenient to define the order of the rotational symmetry $N'$, so that $N' = N$ for $\sigma = +1$ and $N' = N/2$ for $\sigma = -1$.

The essential difference between the potentials stems from their imaginary parts: in the $p\mathcal{PT}$-symmetric potential waveguides are oriented such that local currents from amplifying [bright spots in Fig. 1(b)] to absorbing [dark spots in Fig. 1(b)] domains inside each waveguide are pointed clockwise (such currents are most pronounced inside waveguides, where light intensity is larger, and are much weaker between waveguides), making this azimuthal direction *nonequivalent* to the counterclockwise direction. The profile of such potentials as a function of polar angle $\varphi$ at fixed radius $r$ resembles a one-dimensional $\mathcal{PT}$-symmetric lattice with a *ratchet-like* gain-loss landscape. In the $\mathcal{PT}$-symmetric potential gain and loss domains are exchanged in the neighboring wave-guides [see Fig. 1(c)], hence current directions inside waveguides also alternate making clockwise and counterclockwise directions *equivalent*.

The difference in the discrete symmetries have a profound impact on the spectra of the *linear* eigenmodes $q_m = w_m(\eta,\zeta)\exp(ib^m\xi)$, where $b^m = b^m_{\text{re}} + ib^m_{\text{im}}$ is the eigenvalue and $m \in \mathbb{Z}$ distinguishes the eigenmodes. The basic rotations by $\phi_{\sigma=+1} = 2\pi/N$ or by $\phi_{\sigma=-1} = 4\pi/N$, that leave unchanged the arrangements in each symmetry group, imply the existence of the non-degenerate fundamental state $w_0(r,\phi) = w_0(r,\phi+\phi_\sigma)$ corresponding to a real eigenvalue $b^0$. Higher-order eigenmodes can be represented as angular Bloch waves $w_m(r,\phi) = e^{im\phi}U_m(r,\phi)$, where $U_m(r,\phi) = U_m(r,\phi+\phi_\sigma)$. Their eigenvalues, either real or appearing as complex conjugate pairs, are *always degenerate*. Let us introduce the operator $\mathcal{H} = \mathcal{H}_0 + ip_{\text{im}}\mathcal{R}_{\text{im}}$ in Eq. (1), where $\mathcal{H}_0 = -(1/2)\nabla^2 - p_{\text{re}}\mathcal{R}_{\text{re}}$ is Hermitian. For $\sigma = -1$ the operator $\mathcal{H}$ is $\mathcal{PT}$-symmetric (i.e. $\mathcal{PTH} = \mathcal{HPT}$). If its eigenvalue $b^m$ ($\mathcal{H}w_m = b^m w_m$) is real, then $\mathcal{H}(\mathcal{PT}w_m) = b^m(\mathcal{PT}w_m)$, i.e. $\mathcal{PT}w_m = (-1)^m e^{-im\phi}U_m^*(r,\phi+\pi)$ is also an eigenmode corresponding to the same degenerate real $b^m$. The same is true for the $p\mathcal{PT}$-symmetric operator. For a complex $b^m$ it is the state $\mathcal{PT}w_{N'-m} = (-1)^{N'-m} e^{-i(N'-m)\phi}U_{N'-m}^*(r,\phi+\pi)$ that gives the second eigenmode for the same $b^m$. $\mathcal{H}_0$ also possesses degenerate pairs of eigenvalues $\tilde{b}^m$ ($m \neq 0, N'/2$).

The evolution of two highest pairs of degenerate eigenvalues of $\mathcal{H}$ with increase of the imaginary part of potential can be understood using the simplified model accounting only for those two levels. Let us define eigenmodes $\tilde{w}_m$ and eigenvalues $\tilde{b}^m$ of $\mathcal{H}_0$ ($\mathcal{H}_0\tilde{w}_m = \tilde{b}^m\tilde{w}_m$) and build matrix representation $\langle \tilde{w}_m|\mathcal{H}|\tilde{w}_n\rangle$ for complex operator $\mathcal{H}$. The elements of the resulting matrix are given by $H_{mn} = \tilde{b}^m\delta_{m,n} + ip_{\text{im}}\langle\tilde{w}_m|\mathcal{R}_{\text{im}}|\tilde{w}_n\rangle$. Using $\phi_\sigma$-periodicity of the Bloch modes and potential $\mathcal{R}_{\text{im}}$ one obtains that $\langle\tilde{w}_m|\mathcal{R}_{\text{im}}|\tilde{w}_n\rangle = \int_0^\infty\int_0^{2\pi}\tilde{w}_m^*\mathcal{R}_{\text{im}}\tilde{w}_n r dr d\phi = \delta_{m-n,N'}R_{m,n}$, where the coefficients $R_{m,n}$ depend only on $\mathcal{R}_{\text{im}}$ and $U_{m,n}(r,\phi)$ shapes within the angular interval $\phi \in [0,\phi_\sigma]$. This implies that coupling of linear states with different vorticity $m,n$ is possible with increase of $p_{\text{im}}$ only for $|m-n| = N'$. Thus, in the $p\mathcal{PT}$-symmetric case with $N = N' = 6$ growing imaginary part $ip_{\text{im}}\mathcal{R}_{\text{im}}$ *does not* result in coalescence of the eigenvalues associated with $m = \pm 1$ and $n = \pm 2$. In contrast, in similar $\mathcal{PT}$-symmetric potential with $N' = 3$, coalescence of the eigenvalues is possible. In this approximate model the modified eigenvalues can be found from the matrix

$$H = \begin{bmatrix} \tilde{b}^1 & 0 & 0 & ip_{\text{im}}R_{1,-2} \\ 0 & \tilde{b}^1 & ip_{\text{im}}R_{1,-2}^* & 0 \\ 0 & ip_{\text{im}}R_{1,-2} & \tilde{b}^2 & 0 \\ ip_{\text{im}}R_{1,-2}^* & 0 & 0 & \tilde{b}^2 \end{bmatrix}, \qquad (3)$$

and are given by $b^{1,2} = (\tilde{b}^1 + \tilde{b}^2)/2 \pm [(\tilde{b}^1 - \tilde{b}^2)^2/4 - p_{\text{im}}^2|R_{1,-2}|^2]^{1/2}$. Increasing $p_{\text{im}}$ leads to equality of *two double degenerate* eigenvalues at $p_{\text{im}}^{\text{cr}} = (\tilde{b}^1 - \tilde{b}^2)/2|R_{1,-2}|$, that *remain double degenerate* after that point, but move into the complex plane at $p_{\text{im}} > p_{\text{im}}^{\text{cr}}$. At $p_{\text{im}} = p_{\text{im}}^{\text{cr}}$ the matrix $H$ cannot be represented in diagonal form, but allows representation in the Jordan-block form $H = I \otimes h$, where $I$ is a $2 \times 2$ identity matrix and $h$ is a $2 \times 2$ matrix with elements $h_{11} = h_{22} = (\tilde{b}^1 + \tilde{b}^2)/2$, $h_{12} = 1$ and $h_{21} = 0$. This is an indication of co-existence of two exceptional points at $p_{\text{im}} = p_{\text{im}}^{\text{cr}}$ [31], where *two pairs* of vortex states with $m = \pm 1$ and $m = \pm 2$ simultaneously coalesce.

The numerically calculated spectra, shown in Fig. 2 for potentials with $N = 6$ support the above conclusions. In the $\mathcal{PT}$-symmetric potential one observes coalescence of two *double degenerate* real eigenvalues leading to the appearance of *two double degenerate* complex conjugate eigenvalues at $p_{\text{im}} = p_{\text{im}}^{\text{cr}}$ [we show only eigenvalues that can lead to symmetry breaking]. In contrast, no such coalescence is observed in the $p\mathcal{PT}$-symmetric potential, where a noticeable imaginary parts $b_{\text{im}}^{\pm m}$ do not appear up to the point where the corresponding real parts $b_{\text{re}}^{\pm m}$ approach the edge of the continuous spectrum [Fig. 2(b)]. Such behavior suggests that stable vortex modes may form in the $p\mathcal{PT}$-symmetric potential, where two azimuthal directions are *nonequivalent*, even in the regime where symmetry is already broken in the $\mathcal{PT}$-symmetric case.

Vortex solitons have the form $q = u(\mathbf{r})\exp[i\phi(\mathbf{r}) + ib\xi]$, where $\mathbf{r} = (\eta, \zeta)$, and $u, \phi$ are the field modulus and phase. The latter defines the topological charge $m = (2\pi)^{-1} \oint \nabla\phi d\mathbf{l}$, where the integral is calculated over any closed contour surrounding phase dislocation at $\mathbf{r} = 0$. Eq. (1) then yields $bu = (1/2)\nabla^2 u - \mathbf{j}^2/2u^3 + p_{\text{re}}\mathcal{R}_{\text{re}} u + u^3$ and $\nabla \cdot \mathbf{j} = 2p_{\text{im}}\mathcal{R}_{\text{im}} u^2$, where we introduced the current $\mathbf{j} = u^2 \nabla\phi$. Although the imaginary part of the potential enters only the equation for current, the latter does affect the soliton shape via the first equation. The rigorous simulations reveal that the *charge rule* $|m| \le N/2 - 1$ (for even $N$) established in [13] and connecting maximal topological charge of compact solitons with the order of discrete rotation symmetry of potential, holds in the $p\mathcal{PT}$-symmetric case. Further we consider representative $p\mathcal{PT}$-symmetric potential with $N = 6$ supporting vortex solitons with charges up to $|m| = 2$. We set $p_{\text{re}} = 5$, $a = 0.5$, $\rho = 0.3N$ and use $p_{\text{im}}$ as the main control parameter. Examples of solitons supported by such potentials are depicted in Figs. 3 and 4. All of them feature $N$ pronounced bright spots. Solitons with in-phase and out-of-phase spots and $m = 0$ are presented in Figs. 3(a) and 3(c), respectively. For $p_{\text{im}} \ne 0$ even these simplest $m = 0$ solitons possess nontrivial phase distributions. Solitons with nonzero topological charges are shown in Figs. 3(b), 4(a), and 4(b).

The *central physical result* of this Letter is that the *nonequivalence* of two azimuthal directions in the $p\mathcal{PT}$-symmetric potential causes substantial differences in the shapes and properties of oppositely charged vortex solitons with equal propagation constants. This is in sharp contrast to all previously reported findings on vortex solitons in conservative potentials, where two oppositely charged states are degenerate. The difference is illustrated in Figs. 4(a),(b) for the $m = \pm 2$ states. These are the most stable vortex solitons in the $p\mathcal{PT}$-symmetric structure with $N = 6$. The intensity and local phase modulations are substantially deeper for $m = +2$ than for its $m = -2$ counterpart. The origin of the difference is visible from phase distributions in the second row. Vortex solitons are characterized by the presence of a *global* current associated with vorticity. Such current is counterclockwise for $m > 0$ and clockwise for $m < 0$. Upon stationary propagation net gain/loss experienced by the solitons has to vanish, hence *local* currents directed from amplifying into absorbing domains should appear in each waveguide. The direction of local currents is indicated in Fig. 4 with short gray arrows. While for $m > 0$ the angular directions of all local and global currents are opposite, for $m < 0$ they coincide - see the vector current maps $\mathbf{j}(\mathbf{r})$ in the last row. The difference in directions of local and global currents implies a difference in the current magnitudes $|\mathbf{j}(\mathbf{r})|$ [third row], which, in turn, affect the field modulus distributions and lead to different soliton properties for a fixed $b$. In $\mathcal{PT}$-symmetric potentials the difference between states with opposite charges does not occur because there the directions of local currents *alternate* in neighboring waveguides, so that even if in one waveguide local and global currents have opposite directions, in the neighboring waveguide they coincide [second and fourth rows of Fig. 4(c)]. This leads to an additional azimuthal modulation of the current and field modulus distributions [first and third rows of Fig. 4(c)]. The current modulus distribution now contains two alternating types of spots. The field modulus distribution for $m = -2$ vortex soliton can be obtained from that for $m = +2$ state simply by its rotation by an angle $2\pi/N$, i.e. energy flows of such solitons remain *identical* – a consequence of the *equivalence* of two azimuthal directions in the $\mathcal{PT}$-symmetric potential. The presence of additional azimuthal modulation reduces discrete rotation symmetry of the field modulus distributions in the $\mathcal{PT}$-symmetric structure in comparison with its $p\mathcal{PT}$-symmetric counterpart.

Vortex solitons in the $p\mathcal{PT}$-symmetric structure are characterized by their energy flows $U = \iint |q|^2 d\eta d\zeta$, whose dependencies on $b$ are shown in Fig. 5(a). Solitons exist above a cutoff identical for opposite charges. The energy flow grows with $b$ and saturates at $b \to \infty$. Solitons with negative charges carry *higher* energy flows than solitons with positive charges. The difference $\delta U = U_{m=-2} - U_{m=+2}$ acquires its maximal value for intermediate values of the propagation constants and then gradually vanishes when soliton transforms into $N$ strongly localized almost non-interacting bright spots concentrated within individual waveguides [Fig. 5(b)], consistent with physical expectations. A similar behavior was found for all studied pairs $\pm m$ of nonzero topological charges. The difference in energy flow $\delta U$ between oppositely charged solitons always grows with the imaginary part of potential $p_{\text{im}}$ and it can become comparable with $U$ [Fig. 5(c)]. At

large $p_\text{im}$ values vortex solitons were found to exhibit considerable shape transformations. For example, $N$ spots can fuse into almost a uniform ring at $p_\text{im} \gg p_\text{re}$.

The *nonequivalence* of the azimuthal directions is manifested also in different stability properties of oppositely charged solitons. We found that $m = \pm 2$ vortex solitons are stable in a limited interval of propagation constants $b_\text{cr}^\text{low} \leq b \leq b_\text{cr}^\text{upp}$, but such interval differs substantially for solitons with positive and negative topological charges [compare domains of stability located *between* red dots in Figs. 6(a) and 6(b) and obtained by direct integration of Eq. (1) with perturbed inputs propagated up to $\xi = 10^4$]. The stability domain is nearly three times wider (in terms of $p_\text{im}$) for $m = +2$ solitons than for $m = -2$ ones. For small values of $p_\text{im}$ the lower border of stability domain $b_\text{cr}^\text{low}$ nearly coincides with the cutoff $b_\text{co}$ for existence shown in Fig. 6 by black dots. For $m = +2$ solitons the value $b_\text{cr}^\text{low}$ starts departing from $b_\text{co}$ at $p_\text{im} > 6$ and at one point the lower border of stability domain fuses with upper border $b_\text{cr}^\text{upp}$. Beyond this point solitons become unstable for any $b$. A similar scenario is found for $m = -2$ solitons. The stability domain drastically expands when one approaches the conservative limit $p_\text{im} \to 0$. We also found that $m = 0$ solitons with in-phase spots and $m = \pm 1$ solitons are unstable, while multipole modes, like those depicted in Fig. 3(c), can be stable.

Summarizing, we introduced $p\mathcal{PT}$-symmetric potentials where two azimuthal directions are nonequivalent and uncovered the important implications of such effect in the properties of vortex solitons. Specifically, the properties of the vortex soliton states supported by such potentials were found to depend on both the absolute value of their topological charge and on its *sign*, in complete contrast to all conservative potentials studied so far, where oppositely charged vortex solitons always exhibit equal properties. Note also that the potentials constructed here represent the first example of a $\mathcal{PT}$-*symmetric ratchet* featuring a purely real spectrum in a certain parameter domain.

VVK was supported by FCT (Portugal) under the grants PTDC/FIS-OPT/1918/2012 and UID/FIS/00618/2013.

# Figure captions

Figure 1. (Color online) Profiles of real (a) and imaginary (b) parts of the $p\mathcal{PT}$-symmetric potential, and (c) imaginary part of the $\mathcal{PT}$-symmetric potential with $N=6$. The real parts of potentials are identical. Here and in all contour plots the profiles are shown within the $\eta, \zeta \in [-4.3, +4.3]$ window.

Figure 2. (Color online) Real (black curves) and imaginary (red curves) parts of the eigenvalues of linear eigenmodes of the $\mathcal{PT}$-symmetric (a) and $p\mathcal{PT}$-symmetric (b) structures versus $p_{\text{im}}$ at $p_{\text{re}}=5$. Superscripts $\pm m$ indicate topological charges of beams that can be constructed using linear combinations of corresponding eigenmodes.

Figure 3. (Color online) Field modulus $u$ (first row) and phase $\phi$ (second row) distributions for soliton with in-phase spots (a), for $m=+1$ vortex soliton (b), and for multipole soliton (c) in the $p\mathcal{PT}$-symmetric structure. In all cases $b=1.8$, $p_{\text{re}}=5$, $p_{\text{im}}=10$.

Figure 4. (Color online) Field modulus $u$ (first row), phase $\phi$ (second row), current modulus $|j|$ (third row), and vector current map (fourth row) for vortex solitons with $m=+2$ (a) and $m=-2$ (b) in the $p\mathcal{PT}$-symmetric structure, and for vortex soliton with $m=+2$ (c) in the $\mathcal{PT}$-symmetric structure. In all cases $b=1.8$, $p_{\text{re}}=5$, $p_{\text{im}}=10$. White circle with arrow in the phase distributions indicate the direction of global current associated with vorticity, while short gray arrows indicate the direction of the local currents inside waveguides.

Figure 5. (Color online) (a) Energy flows of $m=\pm 2$ solitons versus $b$ at $p_{\text{im}}=10$. Circles correspond to solitons from Figs. 4(a),(b). (b) Difference $\delta U$ of energy flows of vortex solitons with $m=\pm 2$ versus $b$. (c) Energy flows of $m=\pm 2$ solitons versus $p_{\text{im}}$ at $b=3$.

Figure 6. (Color online) Domains of existence and stability in the plane $(p_{\text{im}}, b)$ for vortex solitons with $m=+2$ (a) and $m=-2$ (b). Solitons exist for $b>b_{\text{co}}$ in the region above black dots and are stable in the interval $b \in [b_{\text{cr}}^{\text{low}}, b_{\text{cr}}^{\text{upp}}]$ between red dots.

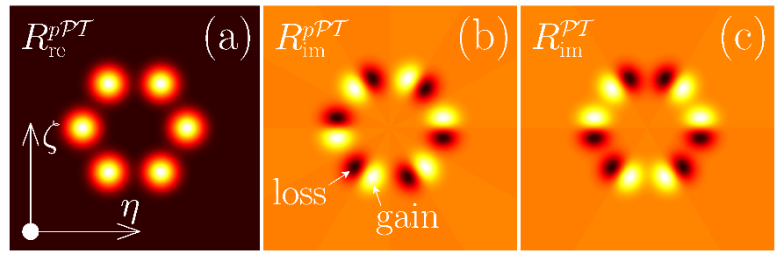

Figure 1. (Color online) Profiles of real (a) and imaginary (b) parts of the $p\mathcal{PT}$-symmetric potential, and (c) imaginary part of the $\mathcal{PT}$-symmetric potential with $N=6$. The real parts of potentials are identical. Here and in all contour plots the profiles are shown within the $\eta, \zeta \in [-4.3, +4.3]$ window.

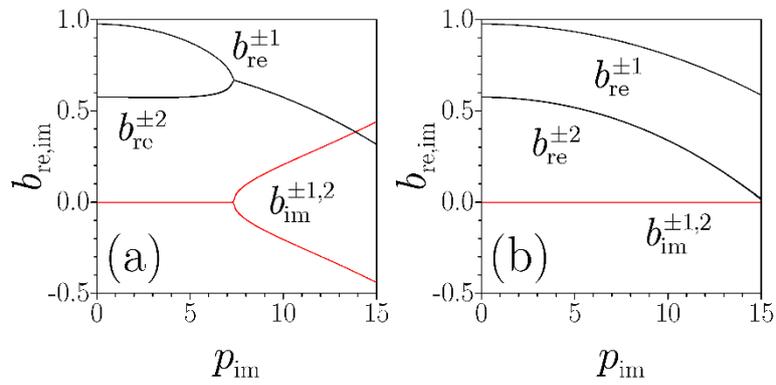

Figure 2. (Color online) Real (black curves) and imaginary (red curves) parts of the eigenvalues of linear eigenmodes of the $\mathcal{PT}$-symmetric (a) and $p\mathcal{PT}$-symmetric (b) structures versus $p_{\mathrm{im}}$ at $p_{\mathrm{re}}=5$. Superscripts $\pm m$ indicate topological charges of beams that can be constructed using linear combinations of corresponding eigenmodes.

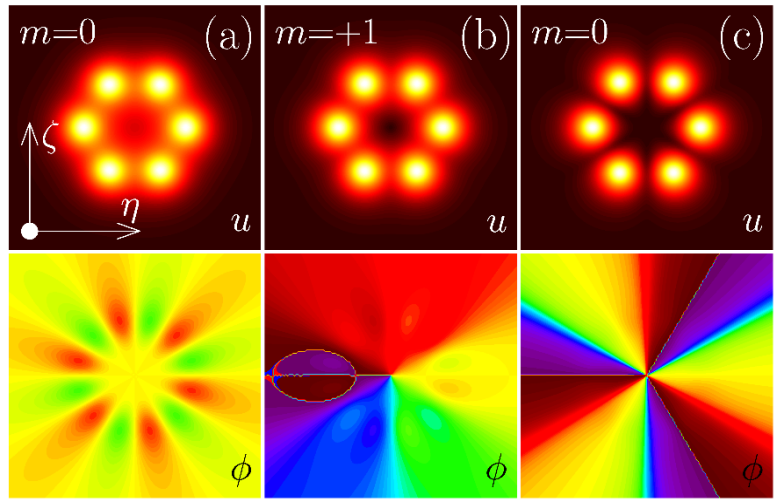

Figure 3. (Color online) Field modulus $u$ (first row) and phase $\phi$ (second row) distributions for soliton with in-phase spots (a), for $m=+1$ vortex soliton (b), and for multipole soliton (c) in the $p\mathcal{PT}$-symmetric structure. In all cases $b=1.8$, $p_{\mathrm{re}}=5$, $p_{\mathrm{im}}=10$.

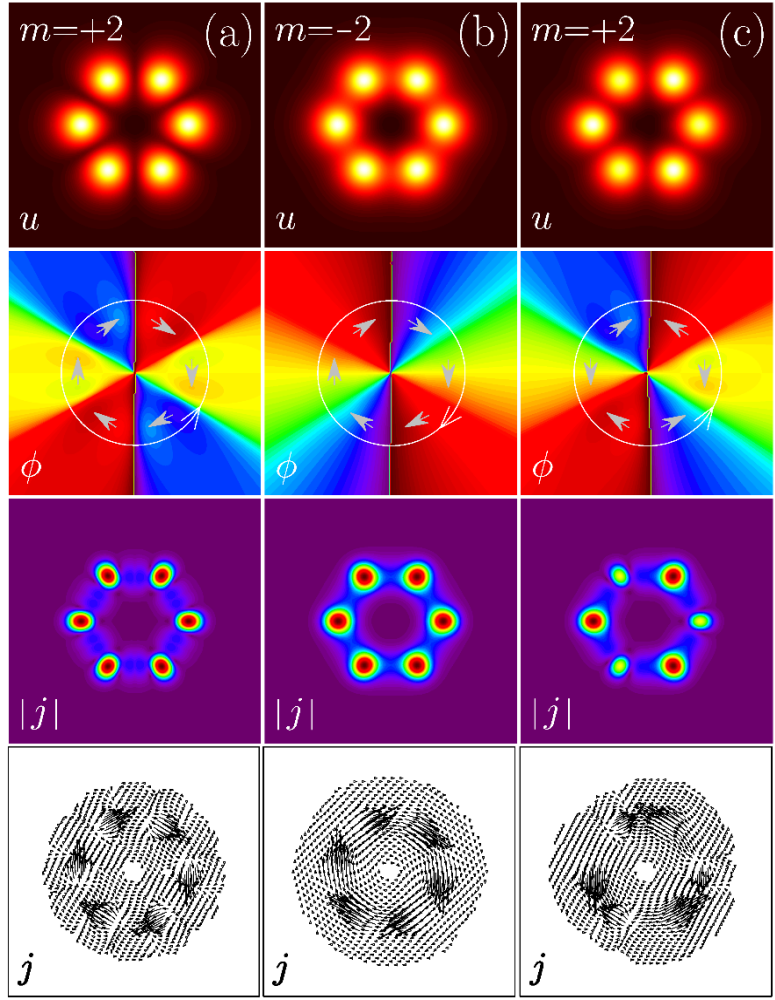

Figure 4. (Color online) Field modulus $u$ (first row), phase $\phi$ (second row), current modulus $|j|$ (third row), and vector current map (fourth row) for vortex solitons with $m=+2$ (a) and $m=-2$ (b) in the $p\mathcal{PT}$-symmetric structure, and for vortex soliton with $m=+2$ (c) in the $\mathcal{PT}$-symmetric structure. In all cases $b=1.8$, $p_{\rm re}=5$, $p_{\rm im}=10$. White circle with arrow in the phase distributions indicate the direction of global current associated with vorticity, while short gray arrows indicate the direction of the local currents inside waveguides.

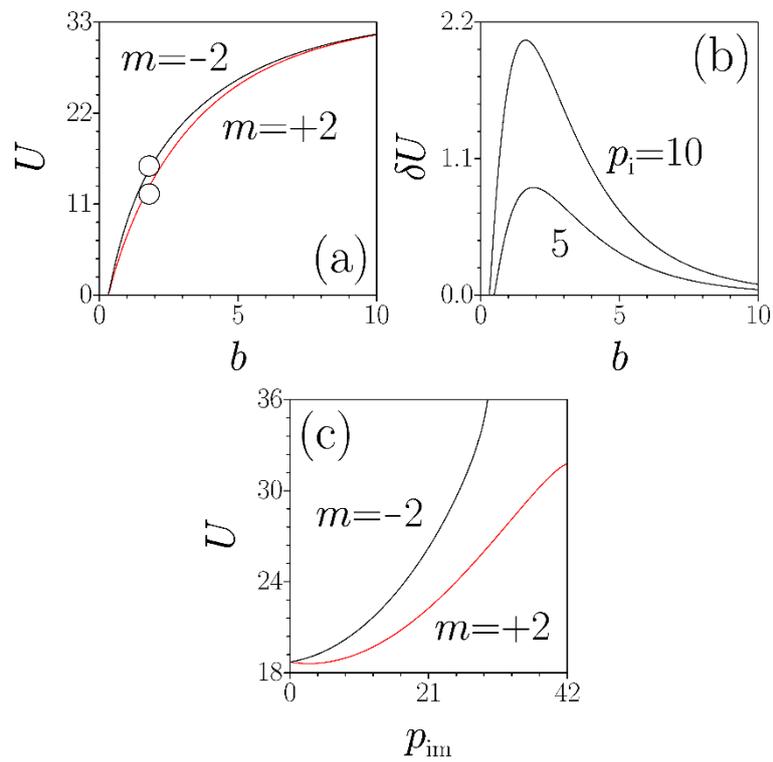

Figure 5. (Color online) (a) Energy flows of $m=\pm 2$ solitons versus $b$ at $p_{\mathrm{im}}=10$. Circles correspond to solitons from Figs. 4(a),(b). (b) Difference $\delta U$ of energy flows of vortex solitons with $m=\pm 2$ versus $b$. (c) Energy flows of $m=\pm 2$ solitons versus $p_{\mathrm{im}}$ at $b=3$.

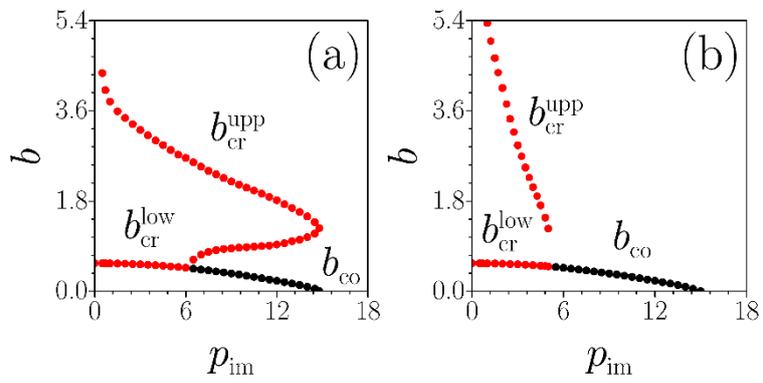

Figure 6. (Color online) Domains of existence and stability in the plane $(p_{\mathrm{im}}, b)$ for vortex solitons with $m = +2$ (a) and $m = -2$ (b). Solitons exist for $b > b_{\mathrm{co}}$ in the region above black dots and are stable in the interval $b \in [b_{\mathrm{cr}}^{\mathrm{low}}, b_{\mathrm{cr}}^{\mathrm{upp}}]$ between red dots.